# INVESTIGATING PERSONALISATION-PRIVACY PARADOX AMONG YOUNG IRISH CONSUMERS: A CASE OF SMART SPEAKERS

Caoimhe O'Maonaigh, Trinity College Dublin, comaonai@tcd.ie

Deepak Saxena, BITS Pilani, deepak.saxena@pilani.bits-pilani.ac.in

**Abstract:** Personalisation refers to the catering of online services to match consumer's interests. In order to provide personalised service, companies gather data on the consumer. In this situation, consumers must navigate a trade-off when they want the benefits of personalised information and services while simultaneously wish to protect themselves from privacy risks. However, despite many individuals claiming that privacy is an essential right to them, they behave contradictorily in online environments by not engaging in privacy-preserving behaviours. This paradox is known as the personalisation-privacy Paradox. The personalisation-privacy paradox has been studied in many different scenarios, ranging from location-based advertising to online shopping. The objective of this study is to investigate the personalisation-privacy paradox in the context of smart speakers. Based on an exploratory study with young Irish consumers, this study suggests a difference between the users and non-users of smart speakers in terms of their perception of privacy risks and corresponding privacy-preserving behaviours. In so doing, it also explains the existence of the personalisation-privacy paradox and offers insights for further research.

**Keywords:** Personalisation, Privacy, Paradox, Smart Speakers.

## 1. INTRODUCTION

Privacy is a concept that our society has long considered a right. However, with the introduction of smartphones and Big Data collection, the attitude towards privacy has arguably been altered. Despite many individuals still claiming that privacy is an essential right to them, they behave contradictorily in online environments by not engaging in privacy-preserving behaviours. This paradox is known as the personalisation-privacy Paradox. The personalisation-privacy paradox has been studied in many different scenarios, such as location-based services, e-commerce, and social media. This study explores the paradox in the context of smart speakers or smart personal assistants. The worldwide number of smart personal assistants' users is expected to grow to 1.8 billion by 2021. Smart speakers, such as Amazon Echo and Google Home, have been becoming more and more popular in recent times with a typical U.S. household having 2.6 smart speakers on average, with 24% of the population owning at least one smart speaker (NPR, 2020).

Against this backdrop, this exploratory study investigates the personalisation-privacy paradox with the users and non-users of smart speakers. It compares the perceptions of a smart speaker user or non-user and investigates how these perceptions differ between the two groups. Hence, the research question for this study is: *How do the users and non-users of smart speakers differ in their privacy perception and privacy-preserving behaviours?*. To answer this question, this paper is organised as follows. Section 2 discusses the privacy risks associated with smart speakers and prior research conducted on personalisation-privacy paradox. Section 3 outlines the methodology in detail. Section 4 discusses the findings of the study, followed by a discussion in section 5. Finally, the last section discusses the theoretical and managerial implications, along with limitations and recommendations for future research.

## 2. LITERATURE REVIEW

### 2.1 Privacy vs Personalisation

Warren and Brandeis (1890) define privacy as 'the right to be alone'. In other words, it allows humans the right to protect their thoughts, property and actions from the knowledge of others. However, with the introduction of Big Data and the Internet of Things, privacy has taken on a new meaning. Information privacy has moved from securing one's personal information to deciding how much information is shared with whom (Smith et al., 1996). It allows consumers to control their information for the ideal representation of self to the outside world (Longo and Saxena, 2020). However, too much information control by individuals runs counter to the personalisation of online services. Personalisation may be defined as delivering appropriate content and services at the right moment to the consumer, consistent with their preferences and past behaviours (Ho and Tam, 2005; Lee and Rha, 2016). In order to provide personalised service, companies gather data from the consumer. Although personalisation benefits are an advantage for consumers, it requires consumers to disclose personal information (Culnan and Armstrong, 1999). In this situation, consumers must navigate a trade-off when they want the benefits of personalised information and services while simultaneously wishing to protect themselves from privacy risks. This trade-off is often termed as personalisation-privacy paradox.

In 2014, the BBC (Wakefiled, 2014) reported that 91% of American consumers think they have lost control over how personal information is collected and used by companies. According to Symantec (2015), 57% of Europeans are worried that their data is not safe in the hands of companies. Interestingly, despite many individuals claiming that privacy is an essential right to them, they behave contradictorily in online environments by not engaging in privacy-preserving behaviours, seemingly resulting in a paradox. For instance, Norberg et al. (2007) demonstrate that despite their stated privacy concerns and intentions, users disclose a considerable amount of personal information. Studies (Acquisti and Grossklags, 2005; Grossklags and Acquisti, 2007) suggest that despite having data concerns, online shoppers are willing to give up their data as long as they receive something in return. This points towards a dichotomy between consumer attitude and consumer behaviour when it comes to privacy concerns. High level of privacy concerns should arguably lead consumers to participate in privacy protection behaviour. However, studies have shown that this concern does not affect consumer privacy behaviour (Kokolakis, 2017).

The personalisation-privacy paradox has been studied in many different scenarios, ranging from location-based services (Xu et al., 2009; Lee and Rha, 2016), e-commerce (Acquisti and Grossklags, 2005; Berendt et al., 2005; Grossklags and Acquisti, 2007; Norberg et al., 2007; Tsai et al., 2011) and social media (Blank et al., 2014; Young and Quan-Haase, 2013). In this regard, it is suggested that privacy behaviour is contextual, and personal data is not a coherent object which is valued the same by each person at all times (Acquisti et al., 2011, 2015; Nissenbaum, 2011). A context-based perspective on the paradox means that privacy concerns differ from person to person and situation to situation. Hence, it is suggested that privacy concerns alone are not adequate predictors of privacy behaviour (Acquisti and Grossklags, 2005) and once has to take contextual factors into account. To investigate the contextual nature of the paradox, this study focuses on the user and non-users of smart speakers. The next section discusses the risk associated with smart speakers and some recent works on privacy perceptions and behaviour.

### 2.2 Smart Speakers and Privacy Perceptions

The two main voice assistants that currently dominate the smart speaker market are Amazon's Alexa and Google Assistant (Marketsandmarkets, 2020). The key to using a smart speaker is its voice recognition function. The smart speaker is 'always listening' until you use a 'wake' word (such as Hey, Alexa or Ok, Google) to wake it up and begin recording what is being said. The recording is sent, via the internet, to a main processing hub. For example, the Amazon speech files are sent to Amazon's Alexa Voice Services (AVS) in the cloud. Here, the voice recognition software deciphers the recording and then sends back the appropriate response to the speaker. It will use the voice command to complete tasks like playing music, answering questions, and controlling lighting and heating etc. While Amazon and Google have both admitted to hiring employees to listen to recordings to improve the speaker's capabilities, it is claimed that all information is treated with the highest of confidentiality (Crist, 2019). Amazon and Google both claim that no personal information is sold to third parties for advertising (Crist, 2019).

Smart speakers are 'always listening', but they are not 'always recording'. However, they can be accidentally activated when they mistake a noise for the command word. There have been instances in the past where a smart speaker has been caught recording everything that an owner has been saying (Russakovskii, 2017). Google has since fixed this issue; however, it highlights how easy it is for the speaker to accidentally start recording without the owner's knowledge. Smart speakers are considered to be relatively secure as they all support WPA-2 encryption and use a secure Wi-Fi connection that prevents data from being hacked from the smart speaker (Panda Security, 2018). However, it is still possible for cyber criminals to control the device, though researchers are currently working on fixing this (Norton, n.d.). The biggest issue is that smart speakers are usually connected to a variety of devices, such as phones and other smart home devices. If a cybercriminal is able to hack into the speaker, they may be able to hack into all your online accounts.

While earlier works mostly focussed on the technical aspect of data security and information privacy, recent studies examine user perception of privacy with smart speakers. Studies suggest that the users have an inadequate grasp of the technology and business ecosystem underpinning smart speakers, resulting in incomplete and inaccurate mental models (Abdi et al., 2019; Huang et al., 2020). Moreover, the users are concerned on the extent of data collection and further sharing with third parties (Cha et al., 2021; Kowalczuk, 2018; Lau et al., 2018; Malkin et al., 2019). Such concerns usually emerge from the lack of clarity on the data collection and sharing practices adopted by companies. In response, users usually adopt from one of the three broad coping strategies – acceptance, mitigation, or avoidance. At one end of the spectrum, some users prioritise convenience over their concerns, and accept the privacy risks (Abdi et al., 2019; Cha et al., 2021; Huang et al., 2020; Lau et al., 2018). Such users are deemed to possess technological optimism (Kowalczuk, 2018). At the other end, the users try to limit their use of smart speakers by avoiding it altogether or intermittently turning it off for some time (Abdi et al., 2019; Huang et al., 2020). Finally, a small group of users try to mitigate privacy risk by making sense of and by trying to use customisable privacy settings available in the device (Cho et al., 2020). However, as Lau et al (2018) report, such privacy controls are often not aligned with the needs of the users, and hence, are not very useful for them.

## 3.     RESEARCH METHODOLOGY

As smart speakers and smart homes are still a relatively new concept, this study seeks to examine the perception of young adults on smart speakers. This research builds upon existing research on the personalisation-privacy paradox but adds novelty by looking at smart speakers from the lens of users and non-users. Therefore, the main research question of this study is: *How do the users and non-users of smart speakers differ in their privacy perception and privacy-preserving behaviours?*.

To answer this question, a qualitative interviewing approach was deemed the most appropriate in understanding the attitudes of young Irish adults. The recruitment of the participants was done using a screening survey. To be an interviewee, a participant had to be a young adult (that is between the ages of 18 and 25 as defined by the Central Statistics Office in Ireland) and be aware of what a smart speaker was. As the aim of the interviews was to see if participants were concerned about privacy and smart speakers without the prompt of the interviewer, the survey was kept very vague to avoid self-selection bias. Options given in the answers were short and to the point. The survey was distributed via social media, email lists and through contacts. The survey allowed participants to submit their contact details to express their interest in being interviewed further and have the chance to win a €10 Amazon voucher. Over 60 potential interviewees participated in the survey. Finally, ten respondents were contacted for the interviews. To balance the perspective, five users and five non-users of smart speakers were then selected for interviewing. Among the ten participants, five were male and five were female. Table 1 provides the details of the participants.

| Participant | Occupation | Gender |
|---|---|---|
| NU1 | Computer Science | M |
| NU2 | Philosophy Student | F |
| NU3 | Unemployed | F |
| NU4 | Physics Student | M |
| NU5 | Unemployed | F |
| U1 | Doctor | F |
| U2 | Sales Manager | F |
| U3 | Technician | M |
| U4 | Sales Associate | M |
| U5 | Radiation Therapist | F |
| *NU = Non-User, U= User.* | | |

**Table 1. Participants in this study**

Due to the time and contextual challenges that presented themselves during the timeline of this research, namely the COVID-19 pandemic, electronic medium was considered the most convenient and effective way to carry out the data collection. In total, eight out of ten interviewees were interviewed via video or audio call, and two were done in person. The interviews took between 20-30 minutes depending on the participant's knowledge and how in-depth their answers were.

Each interview started with a broad question along the lines of "*tell me about smart speakers*", or "*where did you first hear about smart speakers*". An opening set of questions were asked depending on whether they were users or non-users (See Appendix A). Follow-up questions were then asked depending on the answers to previous questions. The transcriptions created by Otter.ai were edited within 24 hours of the interview to fix any mistakes and note any contextual information during the interview sessions. Thematic analysis was conducted to analyse the interviews (Guest et al., 2011). The transcripts were read multiple times in order to get familiar with the content and develop an idea of overarching themes. The data was coded using in-vivo coding in order to keep the participants' own words in the analysis (Saldaña, 2015, Saunders et al., 2016). The next section presents the findings of the analysis.

## 4. FINDINGS

### 4.1 Perception on Privacy and Personalisation

Each participant was asked what privacy meant to them. The element of control is essential to a majority of participants, where they can decide whether or not their information is shared. NU4 used their idea of what privacy meant to them in their decision to not own a smart speaker, noting that with a smart speaker, *"you never really know if your information is where they say it is"*. U4 shared a similar sentiment by expressing that the need for companies to ask for his consent was essential to him in terms of privacy. However, he believes that because the smart speaker has specifically asked for his consent, that he still has control, and his information is kept private. U3 has his whole house set up as a smart home, and yet because he lives alone, he still believes that his life is kept private:

> Well, I live alone. So, pretty much anything that happens in my house I would consider private. Like I know that they say they record all these things, but they don't have people paid to spend hours and hours, listening through them. It would take too many people. So, for the most part, everything I say or do in my house is still private. (U3)

Interestingly, U2 notes that she does not know what privacy means to her due to the age that we are living in.

When asked about personalisation features of the smart speaker and other online activities that result from the surrendering of data, the answers were a mixed bag. Users of the smart speakers favoured the voice recognition, some wishing that it would work a lot better than it does now (U5, U1). When U1 was asked what she would do in order to improve the smart speaker, privacy protection was not a concern for her; however, improved personalisation was. U5 and the majority of the non-users found benefits to only certain types of personalisation. This includes YouTube and Spotify algorithms where they suggest similar artists or videos. U5 believes that there is a benefit to a certain level of personalisation, however, that it can go too far where data is collected that she did not intend to be collected. In a similar manner, U4 notes:

> So, what most people just deem as mainly inconsequential data is usually the stuff you actually find people feel more violated by when other services have it. Just your small day to day things. So, in that sense, by not sharing what I won't ever feel that violation. So, where they're okay with knowing like, okay, regularly searches for traffic to this one location, probably drives to it regularly from the Google Maps information. I'm okay with them figuring out that lives in location A, and works probably at location B, or really like shopping there. I'm okay with that level. (U4)

NU5 expresses a similar sentiment, stating that she is okay with clothes shops online recommending her items because she is in the clothes shop with an intent to buy. However, when she is scrolling through Instagram, and a recommended advertisement for something personal pops up it makes her uncomfortable. NU2 has a different opinion and stated that in the past she has bought items off of Instagram advertisements that were targeted to her based on her history on AliExpress. She finds it *"actually kind of helpful even though it is just them using [her] information and putting up more stuff that they think [she] would like"*.

In contrast, NU3 and NU4 were heavily against the personalisation of online activity, with NU3 going as far as to say that she would sacrifice personalisation for the protection of her data. They both had similar opinions that if they wanted to know about a product or service, they would look it up themselves and do not need it to be suggested to them. NU4 finds that the personalisation features were not of value to him and feels that personalisation of YouTube, Netflix and Spotify accounts actually *"closes you in even, like, in terms of your taste"*. He argues that this results in finding it hard to have a conversation with someone else because they are *"taking in completely different information"* than you are.

### 4.2   Perception on Personalisation vs Privacy Trade-off

Most users find the trade-off of privacy for the features of the speaker to be worth it. U2 finds the speaker 'great', 'convenient' and 'easy'. U4 states that he is happy with the services that he is receiving for his information based on what he is aware that they use it for, as long as nothing comes out that they have breached their terms and services. He believes that he is comfortable with this trade-off as *"all it is doing is taking things you are interested in, and rather than you having to find out about it, it provides that information to you"*. He is also very aware that the company benefits from their data and is comfortable with that as it means better service for him in the long run. U5 is the only user that says that she "begrudgingly accept[s] it" even though she is not overly satisfied with the trade-off.

The majority of the non-users are also aware and are therefore only willing to participate in trade-offs if they perceive some value in the service or product. NU1 states that the perceived trade-off of a smart speaker is not worth the information he would have to give up, whereas other personalised services (e.g., YouTube or Spotify) in his smartphone would be. NU4, who is not willing to give up personal information in any personalisation-privacy trade-off, states that he would be willing to give up his data for the *"greater good of society"*, such as the COVID tracker app, but *"definitely not for* *personalised ads or anything"* as he would *"have no interest in that"*. NU3 is in-between admitting that she participates in the trade-off, not because she wants to but because she does it out of *"laziness or convenience"*. She admits that although she is aware of the personalisation-privacy trade-off, she is unsure why she allows some personalisation on some services and not on a smart speaker.

In this regard, it is important to note that the aspect of control is essential to all participants when it comes to the trade-off. The permission and consent need to be willingly given in order for the participants to feel like they are still in control of their data. U5 notes:

> I feel like it has to be the information willingly given. Okay, is really the crux of it… I will trade my information for the benefit of getting what I'm looking for. But I don't want my information taken from me to push things that I don't want. (U5)

U4 echoes this sentiment by stating that he would like a *"list of what they are using it for, and say, for optimization of [his] experiences, for ads, or other third parties, and like, as long as [he] know[s], [he] can choose to opt-in or opt-out"*.

### 4.3 Privacy-Preserving Behaviour

Finally, it is interesting to see if participants' perception of privacy and privacy protection matched their actions when it comes to their online activity. Despite not having an overall concern for privacy on social media, the majority of users still have their social media accounts set to private. There is a greater fear of people they did not know looking at their profiles and their pictures rather than companies having access to their data (U5, U1, U2). U4 reports that he disabled the voice recognition on his phone as he brings his phone into more private spaces and meetings with his employer. He has the privacy settings on his phone set to collect the barest minimum. All non-users admit to turning off their smart personal assistants on their phone (such as Siri, Google or Bixby). U4 and NU3 mention the VPN as a way to protect their data online. All users admit to allowing cookies on their online activity. In contrast, NU1, NU4 and NU5 do not allow cookies on their website visits.

The majority of users did not use any sort of privacy controls on their speakers. U1, U2 and U5 were not even aware of them. U5 turns off her speaker occasionally when she is having a private conversation but also admits that most of the time she turns it off to save electricity. U3 and U4 are the only users that seem to be aware of any privacy controls on their smart speakers. U3 reported being aware of them but does not use them, only deleting his logs every six months. However, he notes that if he were running a business from his household, he would turn off the smart speakers to avoid business conversations being recorded. U4 is much more privacy-conscious when it comes to his online activity. With regards to his speaker, he makes sure to leave it in the kitchen as he believes the kitchen is not a place where it would hear private details such as bank details or a PPS number. He keeps up to date with any privacy settings updates and has customized the settings so that he is in control of it. As well as this, he states that he would never install a smart speaker in every room, nor would he ever use the smart speaker for his private schedules.

## 5. DISCUSSION

Although all participants are aware of the alleged erosion of privacy in online environments, users and non-users rationalise their beliefs in different ways. Users rationalise their continued use of the smart speaker by weighing up the convenience benefits (Abdi et al., 2019; Cha et al., 2021; Huang et al., 2020; Lau et al., 2018) and noting that their information is everywhere anyway so a smart speaker will not make that much of a difference. Non-users do not feel smart speakers are of enough value to give up their personal information and feel that such devices are an invasion of privacy (Huang et al., 2020; Lau et al., 2018). This is consistent with the privacy calculus model (Kim et al., 2019, Kokolakis, 2017, Gerber et al., 2018) that suggests that consumers partake in a risk-benefit analysis before making a decision. If the anticipated benefits of providing data exceed the perceived worth of their data, the user will willingly disclose their data (Lee and Kwon, 2015; Xu et al., 2009).

Almost all participants mention a trade-off between the benefits of personalised services and the disclosure of their data. Non-users find that some personalised services are worth the risks, such as Spotify, Netflix, YouTube recommendations and the use of a smartphone. This is consistent with recent findings (Cha et al., 2021; Kowalczuk, 2018) that enjoyment facilitates adoption of smart speakers. However, non-users do not find that smart speakers offer worthy enough benefits. Therefore, their motivating factors against the adoption of a smart speaker include privacy concerns and a lack of perceived value that the smart speaker would offer in exchange for their data (Lau et al., 2018). Furthermore, they express that they are not as concerned with a privacy threat (since they are not using the device), they are more so concerned with the principle of their privacy being eroded by big companies and the feeling of surveillance anxiety (Kowalczuk, 2018).

In contrast, users adopt smart speakers primarily for their convenience and entertainment aspects (Cha et al., 2021). Though not as aware of the trade-off, they still mention it in a similar sentiment, stating that the benefits of the smart speaker that they have gained have been worth the collection of their data. There is also some evidence of users' lack of knowledge about the data collection and usage (Abdi et al., 2019; Acquisti and Gross, 2006; Grabowski and Samfelt, 2016; Huang et al., 2020; Malkin et al., 2019). Many users attribute their lack of privacy concerns to lack of knowledge on the topic, trusting the company that owns the speaker and feeling that they are not particularly at risk for a breach in their privacy through the speaker. This supports the construct of privacy as trust (Mourey and Waldman, 2020; Waldman, 2018) meaning that privacy is not seen by the users in terms of data disclosure, but in terms of the trust on the data processor. Interestingly, the privacy controls in smart speakers remain mainly unused by most users, with some not even aware of their existence (Cho et al., 2020; Malkin et al., 2019). This may also be due to the fact that privacy controls reportedly do not fulfil the needs of the users (Lau et al., 2018). The findings also suggest that the users continue to use the service due to privacy fatigue (Choi et al., 2018). Users realise that any other company would also use their data for online advertising and tend to become cynical about their personal data collection (Lau et al., 2018).

## 6.    CONCLUSION

### 6.1    Theoretical Implications

Our findings suggest that there is a difference between the users and non-users in terms of privacy concerns and privacy-preserving behaviour (Gerber et al., 2018; Lau et al., 2018). Privacy concerns lead non-users to participate in far more privacy controls than the users of smart speakers. Non-users are more focused on the morality of privacy invasion, rather than feeling vulnerable for a hack, whereas users seem unfazed and just want to use their speakers. Even though all participants feel that privacy is in having control over your data, users feel that they still have control as they consented to their data being used. Non-users feel that a smart speaker would take that control away. Perhaps for this reason, non-users engage in more privacy control measures that the users do.

The distinction between the users and non-users might help explain the personalisation-privacy paradox in the sense that while the non-users express privacy concerns and engage in privacy protection, it does not translate to privacy protection behaviour among the users who wish to enjoy greater benefits of personalisation. In this sense, the paradox is formed not due to inconsistent behaviour of the users but because the studies combine the data from the users and that from the non-users in their analysis. Highlighting the contextual nature of privacy, this calls for the studies on the personalisation-privacy paradox differentiating between the users and non-users of the technology.

### 6.2    Managerial Implications

In order for companies that own smart speakers to broaden their customer base, they need to address the privacy concerns that many non-users have. Understanding the concerns of non-users is important for companies to turn them into users. Existing research suggests that the users' attitudes towards the service improves when privacy controls are supported by the service provider (Tucker, 2014; Mourey and Waldman, 2020), and complemented with content controls (Cho et al., 2020). As our findings suggest, this may not result in the consumers actually making use of tighter controls, yet it may help in building trust on the company and its services.

Companies need consumer data in order to optimize services and personalise service to their customers. However, being transparent with this information is important to consumers. Companies should consider being clearer in their terms of conditions and not use too much "technical jargon" (U4) in order to draw in more customers and develop a level of trust. Disclosures on data collection are shown to improve trust in the service provider (Aguirre et al., 2015). Service providers need to clearly disclose what data is collected, how long it is stored for, how it is protected, why exactly this data is being collected, and who has access to this data (Malkin et al., 2019). Recently enacted General Data Protection Regulations (GDPR) may provide a suitable guideline in this regard. Understanding and catering for consumers privacy concerns will help companies expand their customer base. Educating consumers on the trade-off between data and services will make consumers feel in control of their data, thus increasing the chances of them having a positive experience with their smart speaker.

### 6.3 Limitations and Future Research

This study should be seen as the first step towards understanding the complexity of the personalisation-privacy paradox in relation to smart speakers. However, a number of limitations of this study are to be highlighted. Due to time and resource constraints, the sample for this study was rather small, therefore limiting the generalisability of this study. All participants were Irish between the ages of 18 and 25, limiting the diversity of the sample. Due to purposive sampling, the interviewees were selected based on their relevance to the study and were not randomly selected, thus resulting in a potential bias. Finally, since the results are qualitative in nature, although offering valuable insights into the research of the paradox, they do not offer definitive results that represent the population.

This study is one of the first to look at the paradox through the lens of the smart speaker and therefore can be built upon. First and foremost, the results need to be generalised by conducting a large-scale quantitative study on the differences among the users and non-users of smart speakers. Second, illustrating the contextual nature of privacy, this study primarily refers to the attribute of being a user or a non-user. Further research could expand on this list of characteristics to include variables such as age, gender, cultural background, and technological acumen.

## REFERENCES AND CITATIONS

# 7.     Appendix A: Interview Opening Questions

**Users**
- Which smart speaker do you have?
- Why did you decide to get that one?
- So, what is your overall perception of smart speakers?
- Where do you keep your smart speaker? Why?
- When do you first hear about smart speakers?
- Why did you get a smart speaker?
- How has your experience been with the speaker so far?
- Would you be able to explain how a smart speaker works to someone who has never heard of a smart speaker?

**Non-users**
- What is your overall perception of smart speakers?
- When do you first hear about smart speakers?
- Why do you not own a smart speaker? Have you ever considered buying a smart speaker?
- Have you ever interacted with a smart speaker?
- Would you be able to explain how a smart speaker works to someone who has never heard of a smart speaker?